\documentclass[twocolumn,prl,groupedaddress]{revtex4}

\usepackage{amsmath,amsfonts}
\usepackage{graphicx,multirow}
\usepackage{epsfig}
\usepackage{slashed}
\usepackage{psfrag}
\usepackage{color}
\PassOptionsToPackage{caption=false}{subfig}
\usepackage{subfig}
\usepackage{booktabs}
\usepackage{hyperref}
\usepackage{ulem}

\newcommand{\be}{\begin{equation}}
\newcommand{\ee}{\end{equation}}

\newcommand{\GeV}{\text{ GeV}}

\newcommand{\fig}[1]{Fig.~\ref{fig:#1}}

\begin{document}

\title{Tagging Partially Reconstructed Objects with Jet Substructure}

%

%
\author{Marat Freytsis}
\affiliation{Department of Physics, Harvard University, Cambridge MA, 02138, USA}
\author{Tomer Volansky}
\affiliation{Raymond and Beverly Sackler School of Physics and Astronomy,\\ Tel-Aviv University, Tel-Aviv 69978, Israel}
\author{Jonathan R. Walsh}
\affiliation{Ernest Orlando Lawrence Berkeley National Laboratory, \\ University of California, Berkeley, CA 94720, USA}
\affiliation{Berkeley Center for Theoretical Physics, \\ University of California, Berkeley, CA 94720, USA}

\begin{abstract}
We present a new tagger which aims at identifying partially reconstructed objects, in which only some of the constituents are collected in a single jet. As an example, we focus on top decays in which either part of the hadronically decaying $W$ or the $b$ jet is soft or falls outside of the top jet cone. We construct an observable to identify remnant substructure from the decay and employ aggressive jet grooming to reject QCD backgrounds. The tagger is complementary to existing ones and works well in the intermediate boost regime where jet substructure techniques usually fail. It is anticipated that a similar tagger can be used to identify non-QCD hadronic jets, such as those expected from hidden valleys. 
\end{abstract}

\maketitle


\section{Introduction}
\label{sec:intro}

The successful discovery of the Higgs boson at the Large Hadron Collider (LHC)~\cite{Aad:2012tfa,Chatrchyan:2012ufa} together with null results in new physics searches stresses the need to improve and develop more sophisticated techniques to search for rare or exotic phenomena. Considerations related to the hierarchy problem typically predict new physics which couples to the top quark and which is often charged under QCD. As a consequence, such new physics may be buried under immense hadronic background.

Interesting classes of models that may escape detection are ones  that admit partially reconstructed objects.   These occur when, for example, constituents from the decay or shower of a heavy particle are soft, not confined to a single jet, or escape detection altogether.  Two  examples are hadronic tops in which either part of the hadronically decaying $W$ or the $b$-quark falls outside of the top-jet cone, and hidden valleys, in which some of the hidden-sector particles are stable and escape detection while the others decay to hadrons. In both cases, some of the information on the hard parton which initiates the jet is missing and the jet no longer has an obvious hard scale to distinguish it from QCD. As a consequence it is hard to identify such events and with existing tools new physics of this kind will go unnoticed.

To improve on the situation, a better understanding of the substructure of such jets is needed. In recent years, significant progress has been made in developing tools for studying the substructure of hadronic events in order to disentangle {new physics} signals from background (see Ref.~\cite{Abdesselam:2010pt,Altheimer:2012mn,Altheimer:2013yza} and references therein). Most existing techniques are effective when used on boosted objects, and are therefore useful at discovering heavy new physics particles that decay hadronically. On the other hand, partially reconstructed jets are best studied in the intermediate regime of moderate boost. The  production rates in this regime are often enhanced by many orders of magnitude and hence may play a crucial role in discovering new physics.   The available tools, however, have significantly reduced sensitivity in this region of parameter space~\cite{Plehn:2013paa} despite its discovery potential.  This letter focuses on this moderate-boost region, demonstrating its utility by studying partially reconstructed tops.  The case of hidden valleys will be presented elsewhere.

\begin{figure*}[t]{
\begin{center}
\includegraphics[width=.49\textwidth]{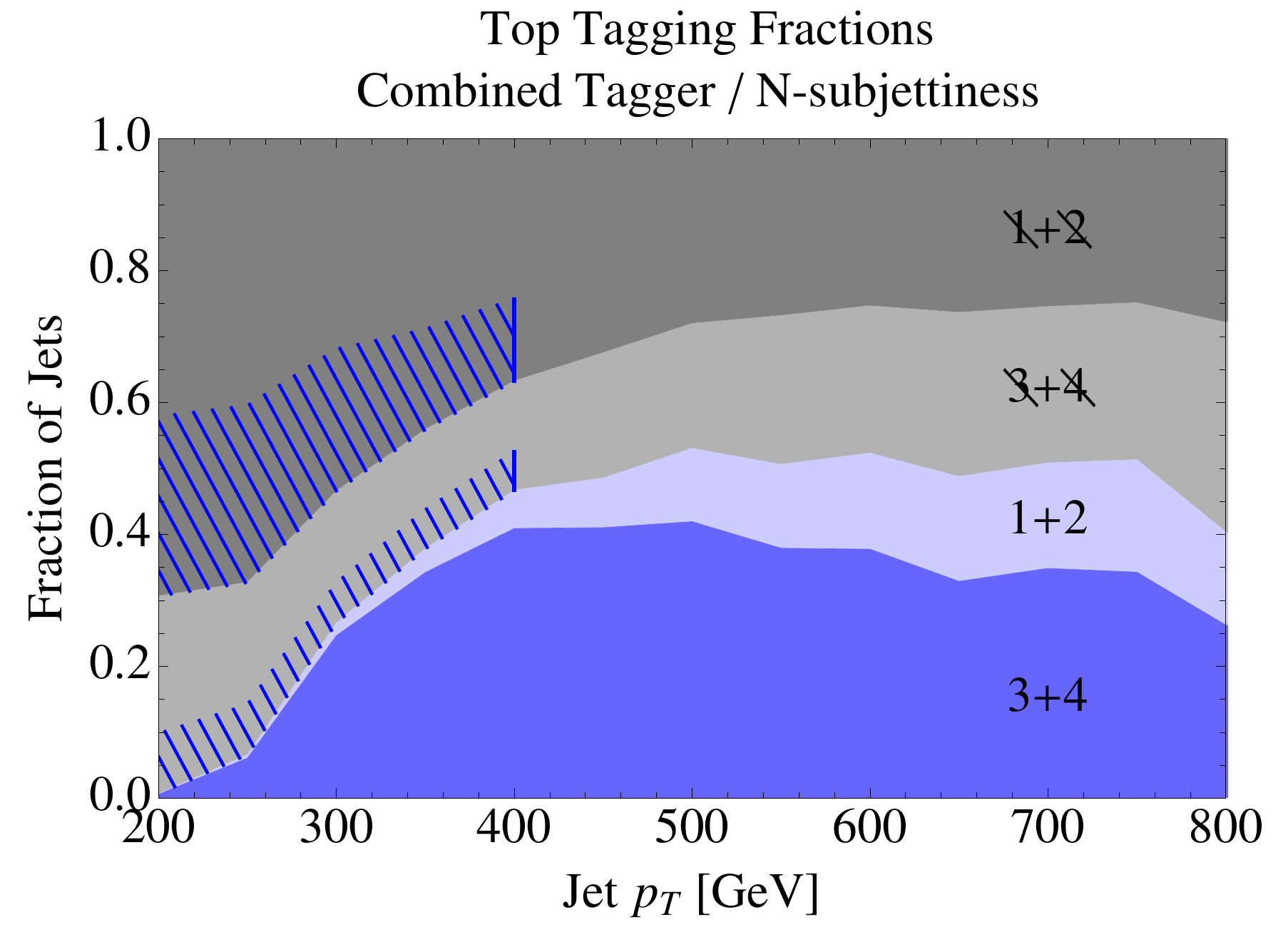} 
\includegraphics[width=.49\textwidth]{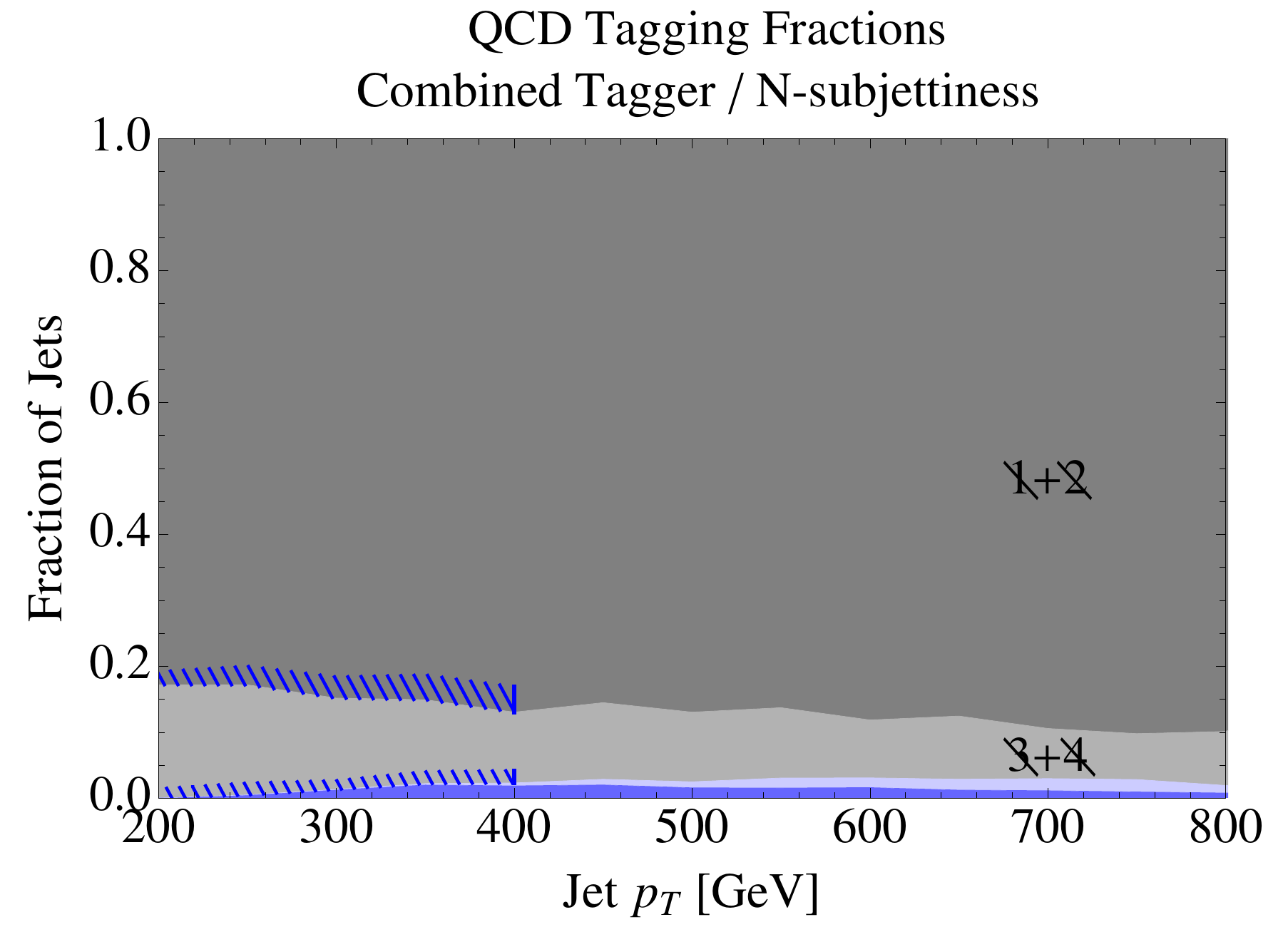} 
\end{center} 
\vspace{-1em}
{ \caption[1]{Subjet fractions for top ({\bf left}) and QCD ({\bf right}) jets as a function of jet $p_T$.  The number of subjets is calculated using the Johns Hopkins top tagger~\cite{Kaplan:2008ie}, and we group the subjets into two categories: 1 or 2 (``1+2'') subjets and 3 or 4 (``3+4'') subjets.  The shaded regions correspond to the $N$-subjettiness top tagger, with the tagged fractions shown in blue and the untagged in gray.  For $p_T < 400$ GeV, we show, hatched in blue, the additional fraction of untagged jets from $N$-subjettiness that are tagged by our top tagger.  }
\label{fig:subjetfracs}} }
\end{figure*}

Top quarks are a standard testing ground for boosted techniques and are considered a standard candle for comparison between methods \cite{Abdesselam:2010pt,Altheimer:2012mn}.    The LHC experiments have used a variety of techniques to identify boosted top quarks, and used them to search for new physics signals~\cite{CMS:2011bqa,Aad:2012dpa,Aad:2012raa,Chatrchyan:2012ku,TheATLAScollaboration:2013qia,Aad:2013gja,CMS:2014aka}.    In the case of fully reconstructed boosted tops, the jets typically admit 3 subjets and counting them  provides a powerful discrimination against QCD background. However, when only two of the top decay products are reconstructed as clear subjets, the resulting substructure exhibits only 2 subjets and does not reconstruct the top. This sample is significantly more difficult to separate from background since there is no resonance mass to cut on, and due to the large QCD background in which a hard gluon splitting yields 2 subjets.   

Despite the above, we argue below that partially reconstructed tops may be efficiently identified. Our key observation is that such jets have radiation patterns that differ from a vast majority of QCD jets, even though the partially reconstructed tops may not have kinematic invariants on which to base selection.  We introduce an observable that is sensitive to these differences, and utilize aggressive jet grooming to further remove QCD jets with multiple subjets which may fake a partially reconstructed top.  The combination results in a robust top tagging tool which works well in the intermediate-boost regime and that is complementary to existing techniques. While we focus on top quarks as a case study, the variables we consider are useful beyond this example, and may be used to identify other partially reconstructed objects in the intermediate boosted regime.

\section{Kinematics Properties of Unreconstructed Decays}
\label{sec:kinematics}

The tagging of boosted heavy particles is a balance between two competing effects.  On the one hand, the direct production cross-section of heavy  objects (which decay into colored final states) falls with increasing momentum, while on the other, tagging with substructure techniques becomes more efficient at higher boost. Consequently, there is often a kinematic regime at intermediate boost where the cross section is large compared to the boosted regime but where most taggers do not effectively function~\cite{Plehn:2013paa}.  These unreconstructed decays can be targeted with some simple considerations about their kinematics, as we now explain.

When a boosted particle such as a $Z$ undergoes a $1\to2$ hadronic decay,  some fraction of those decays can be reconstructed in a single jet. The decays can be characterized by the angle $\theta$ between the forward decay product and the boost direction in the rest frame of the $Z$.  When $\theta \sim \pi/2$, the two quarks from the decay will have similar momenta in the lab frame and the subsequent hadronic final states can be reconstructed in a fat jet. For $\theta\sim 0$, the decay products become asymmetric, with one  softer than the other, and further apart. As a consequence, the soft particle is emitted away from the jet direction, which lies along the hard decay product, and reconstruction is challenging.  Thus, while symmetric decays can be reconstructed in a single jet, asymmetric ones may not be.  Furthermore, these asymmetric decays mimic QCD-like topologies with a hierarchy of parton energies, and so the ability to differentiate these already challenging decays from QCD backgrounds degrades.  Thus, the efficiency to tag boosted $Z$ bosons significantly decreases in this region of phase space.   

Consider now the more complex case of a top quark which undergoes two successive $1\to2$ decays.  Two main parameters are of interest to understand the top reconstruction efficiency: $\theta_W$, the angle of the $W$ to the top boost direction in the top rest frame, and $\theta_q$, the angle of the forward quark from the $W$ decay (defined in the $W$ rest frame) to the $W$ boost direction in the top rest frame.  When $\theta_W \sim 0$, the $W$ will carry most of the top quark's momentum and the $b$ quark will be soft and at a wide angle.  Such a decay will lead to the boosted $W$ being easily reconstructed, but the top will not be.  When $\theta_W \sim \pi$, the $W$ has a small boost in the lab frame and hence the two quarks from the $W$ decay are not reconstructed in the same jet.  Similarly, if $\theta_q \sim 0$, then the $W$ decay products are asymmetric in energy and at wide angles, also leading to an unreconstructed top.  It is only when $\theta_W$ and $\theta_q$ are both not extremal that the top has a large reconstruction efficiency.  Furthermore, when the top is not reconstructed, it is often one of the quarks from the $W$ decay that is lost, meaning the $b$ and the other quark are clustered in the same jet, which is a configuration without any resonance mass to select on.  

The above considerations hold in both the intermediate- and highly-boosted regimes.  
As an illustration, in \fig{subjetfracs} we show the fraction of tagged and untagged tops ({\it left}) and the fraction of mistagged and vetoed QCD jets ({\it right}), using the N-subjettiness top tagger~\cite{Thaler:2010tr,Thaler:2011gf} operating at the 50\% efficiency point (as computed over a range in jet $p_T$).  Each sample is further divided according to the number of subjets (as defined by the Johns Hopkins (JHU) tagger~\cite{Kaplan:2008ie}) found in the event.  We note that the fraction of 4 subjets is numerically small for both tops and QCD, and top jets additionally only rarely have a single subjet.  We see that as the jet $p_T$ decreases, the rate of untagged jets significantly increases, and a non-negligible fraction of untagged jets with ${\le}2$-subjets persists.   
The hashed regions for $p_T < 400$ GeV in each plot indicate the fraction of untagged jets (by $N$-subjettiness) that are tagged by the tagger we describe below.

From the above discussion we identify two interesting types of unreconstructed tops in which not all constituents are found in a single jet: the ones in which the $b$ jet is soft and the ones in which the $W$ is soft.  In the former, the $W$ is boosted into a single jet but the $b$ is at a wide angle, while in the latter we typically find that the $b$ is clustered with one of the quarks from the $W$ decay and the other quark is at a wide angle.  In both cases, one expects roughly equal energy sharing between the quarks in the same jet, which is uncharacteristic of QCD jets that preferentially have soft splittings.  Therefore, a useful kinematic handle is the intrinsic two-prong nature of unreconstructed top jets, in which neither of the two subjets are soft.  We will focus our tagger around observables and techniques that exploit this feature.

A variety of different methodologies have been adopted in order to tag the hadronic decays of boosted tops~\cite{Kaplan:2008ie,Thaler:2008ju,Almeida:2008yp,Almeida:2008tp,Ellis:2009su,Plehn:2009rk,Ellis:2009me,Plehn:2010st,Almeida:2010pa,Thaler:2010tr,Hook:2011cq,Thaler:2011gf,Jankowiak:2011qa,Soper:2012pb,Larkoski:2013eya,Schaetzel:2013vka,Larkoski:2014zma}.  All of these search strategies are focused on tops where the decay may be reconstructed in a single jet, typically by identifying 3 well-separated subjets.  Although combinations of techniques show improved tagging power, they are often limited by the fact that the subsets of tops being tagged are significantly overlapping (to a greater degree than the QCD jets that fake these tops).  However, the tagger discussed below targets a kinematically distinct sample of tops and therefore is largely orthogonal to other techniques, meaning that the combination with other taggers shows increased power in top tagging.

\section{Tagging Partially Reconstructed Objects}
\label{sec:tagger}

The common currency for jet substructure is subjets: isolated clusters of high-$p_T$ radiation in the jet.  Most substructure tagging methods count the number of subjets in a jet (sometimes subject to some additional cuts per subjet) and select those matching the expected number from the boosted decay, such as 3 for the top quark (or 4 if allowing a hard radiated gluon).  Techniques to count subjets typically use declustering algorithms (e.g.,~\cite{Butterworth:2008iy,Kaplan:2008ie,Plehn:2009rk,Hedri:2013pvl}) or shape measurements (e.g.,~\cite{Thaler:2010tr,Thaler:2011gf,Jankowiak:2011qa,Larkoski:2014zma}).  More general jet grooming methods, such as pruning and trimming~\cite{Ellis:2009su,Ellis:2009me,Krohn:2009th} can also be used to shape the jets, but must be supplemented with a subjet counting method like those above. In top tagging, present approaches explicitly remove cases where the top is not fully reconstructed, {i.e.,} when there are fewer than 3 explicitly identified subjets.

Of course, having 3 subjets does not guarantee tagging. In the intermediate $p_T$ regime, 3- and 4-subjet events are often misidentified and do not correspond to the actual top decay products. Consequently, additional (typically mass) cuts applied to combinations of subjets in existing top-taggers are designed to veto such events. Conversely, in the boosted regime, as the $p_T$ is raised, the subjets increasingly correspond to the true decay products, but these additional cuts are still necessary to control QCD backgrounds. The above is illustrated in \fig{subjetfracs} where it is apparent that a sizable sample of top jets, with any number of subjets, is not tagged, with the situation significantly worse at lower $p_T$.

The partially reconstructed tops we target in this paper are precisely those tops missed by conventional top taggers. This sample is dominated by events where either the $b$ quark or one of the quarks from the $W$ is soft or escapes the jet. In the latter case, due to the sequential nature of the top decay, the phase space ensures that such configurations are approximately 3 times more likely to contain a $b$ quark and one of the quarks from the decay of the $W$ rather than a $W$ alone. To tag such events, it is best to assume as little about the precise substructure of the jet as possible.

\begin{figure*}[t!]{
\begin{center}
\includegraphics[width=.48\textwidth]{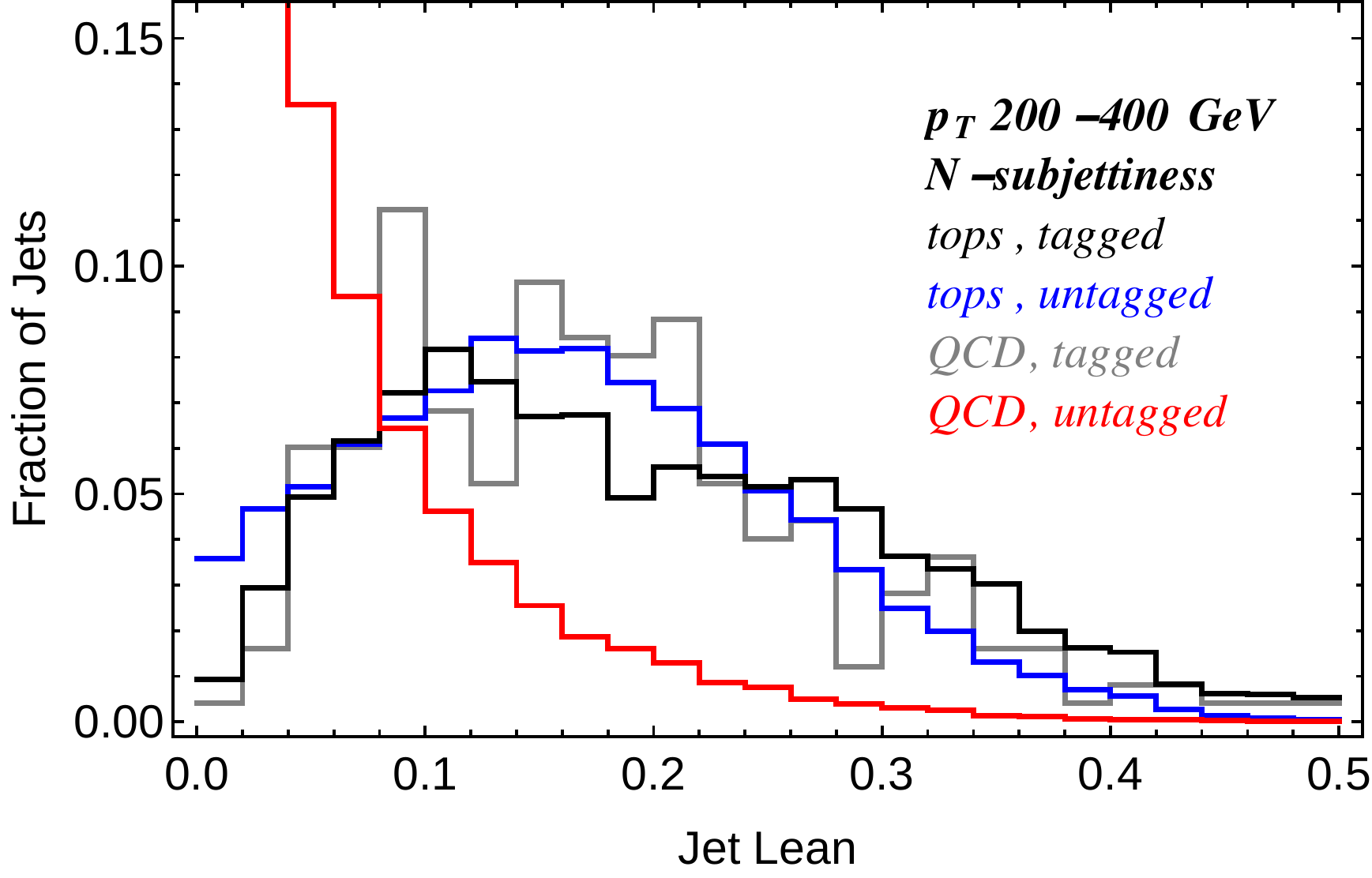} 
\includegraphics[width=.48\textwidth]{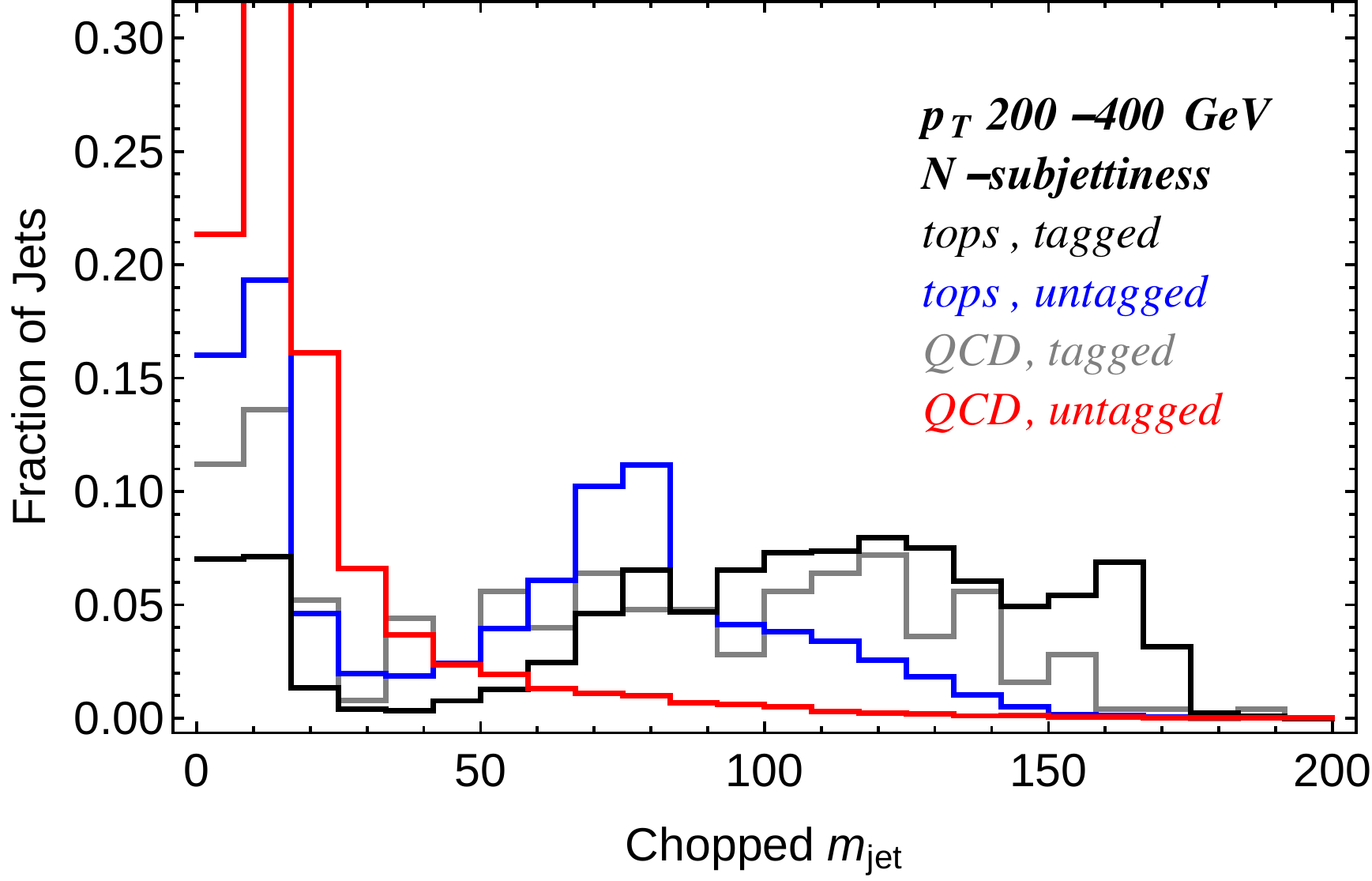} 
\end{center} 
\vspace{-1em}
{ \caption[1]{Distributions of variables used in our tagger: jet lean ({\bf left}) and the chopped jet mass ({\bf right}). These variables are shown for samples of tagged and untagged top and QCD jets.}
\label{fig:dRtau1andmtrim}} }
\end{figure*}

Our first variable is therefore one that allows us to remove the QCD sample with no hard splittings while making no assumptions about the number of hard subjets in the signal sample. We consider 1-subjettiness with variable angular weighting exponent $\beta$~\cite{Thaler:2010tr},
\begin{equation}
  \label{eq:tau1def}
  \tau_1^{(\beta)} = \frac{1}{p_{T,J}} \min_{\hat{n}} \sum_{i \in \text{jet}} p_{T,i} \Delta R_{n,i}^\beta\,.
\end{equation}
For small angles, this sum can be written as $\sum_i z_i \theta_i^\beta$, with angles defined with respect to the axis that minimizes the variable, and $z_i$ the fraction of the jet $p_T$ carried by the $i$th particle. As detailed in Ref.~\cite{Larkoski:2014uqa}, for $\beta = 1$ this sum is always minimized when the axis is aligned with the most energetic particle and if particle splittings are strongly ordered in angle. Alternatively, for $\beta = 2$, this is minimized when the axis is aligned with the jet axis itself in the narrow jet limit.

In the decay of a boosted object, if at least two subjets are reconstructable, the $\tau_1^{(1)}$ axis precisely tracks the hardest subjet, while the $\tau_1^{(2)}$ axis continues to fall on the jet axis. The two axes are then expected to be $O(1)$ apart. Conversely, in the case of a QCD jet with no hard splittings, the jet axis can only move away from the hardest particle when that particle undergoes a collinear splitting. This distance is expected to only be $O(\theta_c)$ from the hardest particle in the jet, with $\theta_c$ the angle of the widest collinear splitting in the shower. We then expect that this distance to still be $O(\theta_c) \approx m_J/p_T \ll 1$. All of this remains true for an arbitrary number of subjets.   

This suggests that a cut on a minimal separation between the $\tau_1^{(1)}$ and $\tau_1^{(2)}$ axes will be able to  eliminate QCD jets with no hard spittings in its parton shower, while remaining agnostic about the nature of the signal. We call this variable the {\it jet lean}, or $\Delta R_{\tau_1}$, the distance between the $\tau_1^{(1)}$ and $\tau_1^{(2)}$ axes. (As noted above, the $\tau_1^{(2)}$ axis and jet axis can be interchanged with no loss of performance.) The discriminating power of jet lean is illustrated on the left of \fig{dRtau1andmtrim}, which shows its distribution for the tagged and untagged top and QCD samples. A clear separation is visible. Demanding that jet lean be above a certain value will eliminate the QCD background coming from jets that get their mass from a wide, diffuse parton shower with no discernible substructure.

The remaining QCD sample with 2 or more subjets can be reduced further, but as we desire to remain agnostic as to the number of reconstructed subjets in the signal sample, we need to exploit the kinematic properties of the subjet splittings. Coming from on-shell decays, top decay products tend to have fairly democratic energy sharing among subjets, while QCD jets, dominated by singularities of the splitting functions even in the case of hard splittings, tend to produce a hierarchy in subjet $p_T$.  While one can check for this condition by simply declustering a jet once, and looking at the $p_T$ ratio of the two subjets, superior discrimination is achieved using jet grooming algorithms~\cite{Ellis:2009me,Krohn:2009th}. This allows us to more efficiently isolate the leading hard subjet.  Because of the presence of soft radiation in the jet from the unidentified components of the top decay, grooming the subjets before measuring their $p_T$ can be more effective than it otherwise would be. 

Here we use trimming, which reclusters the jet into subjets  of small radius $R_\text{sub}$. Only subjets with a $p_T$ fraction of the jet $p_T$ greater than a parameter $f_\text{cut}$ are kept. By running the trimming procedure with a much larger value of $f_\text{cut}$ $(\approx 0.20)$ than typically used, one may veto softer subjets if their $p_T$ fractions are small enough, instead of simply removing soft radiation which is the usual goal of trimming. We call this aggressive use of trimming, {\it chopping}.  Chopping then preferentially removes softer subjets from QCD and converts some of the 2-subjet QCD sample into 1-subjet jets. A cut on the chopped jet mass will then remove these jets.  The behavior of this variable  is illustrated in the right plot of \fig{dRtau1andmtrim}, where the distribution of the chopped jet mass is shown for the tagged and untagged top and QCD samples.

To summarize, our tagger consists of a two-step procedure with parameter values, 
\begin{equation}
  \label{eq:ourcuts}
  \Delta R_{\tau_1} \gtrsim O(0.1)\,, \qquad \text{chopped }m_J \gtrsim O(50\GeV)\,,
\end{equation}
giving good separation between boosted top and QCD jets.   
This combination allows us to simultaneously veto massive QCD jets with no discernible substructure and purify the remaining sample of jets with hard splittings that are more QCD-like, without imposing a particular subjet counting or mass window on the signal.

\section{Efficiency Studies and Results}
\label{sec:efficiency}

\begin{figure*}[t!]{
  \begin{center}
    \includegraphics[width=.32\textwidth]{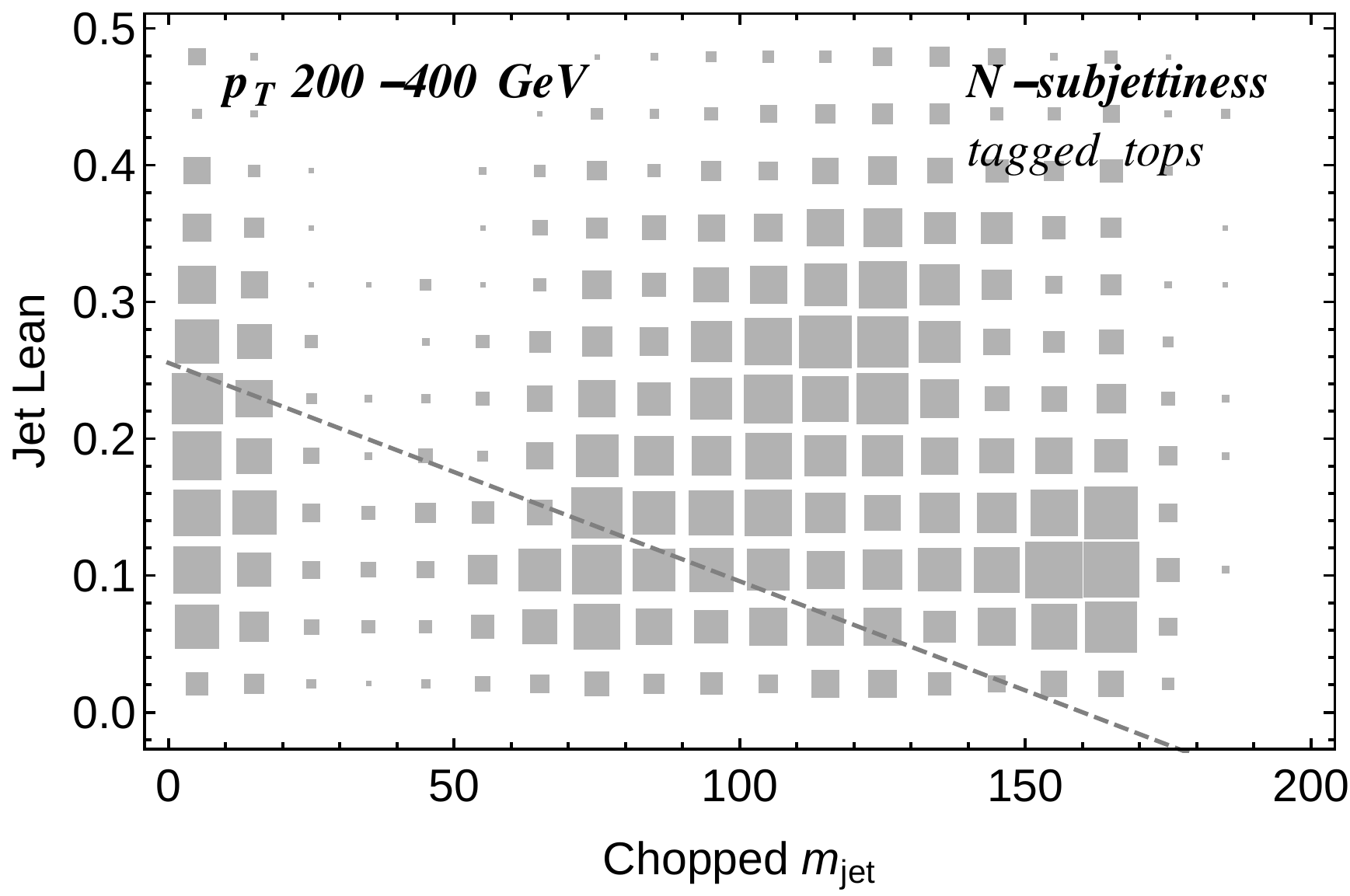} 
    \includegraphics[width=.32\textwidth]{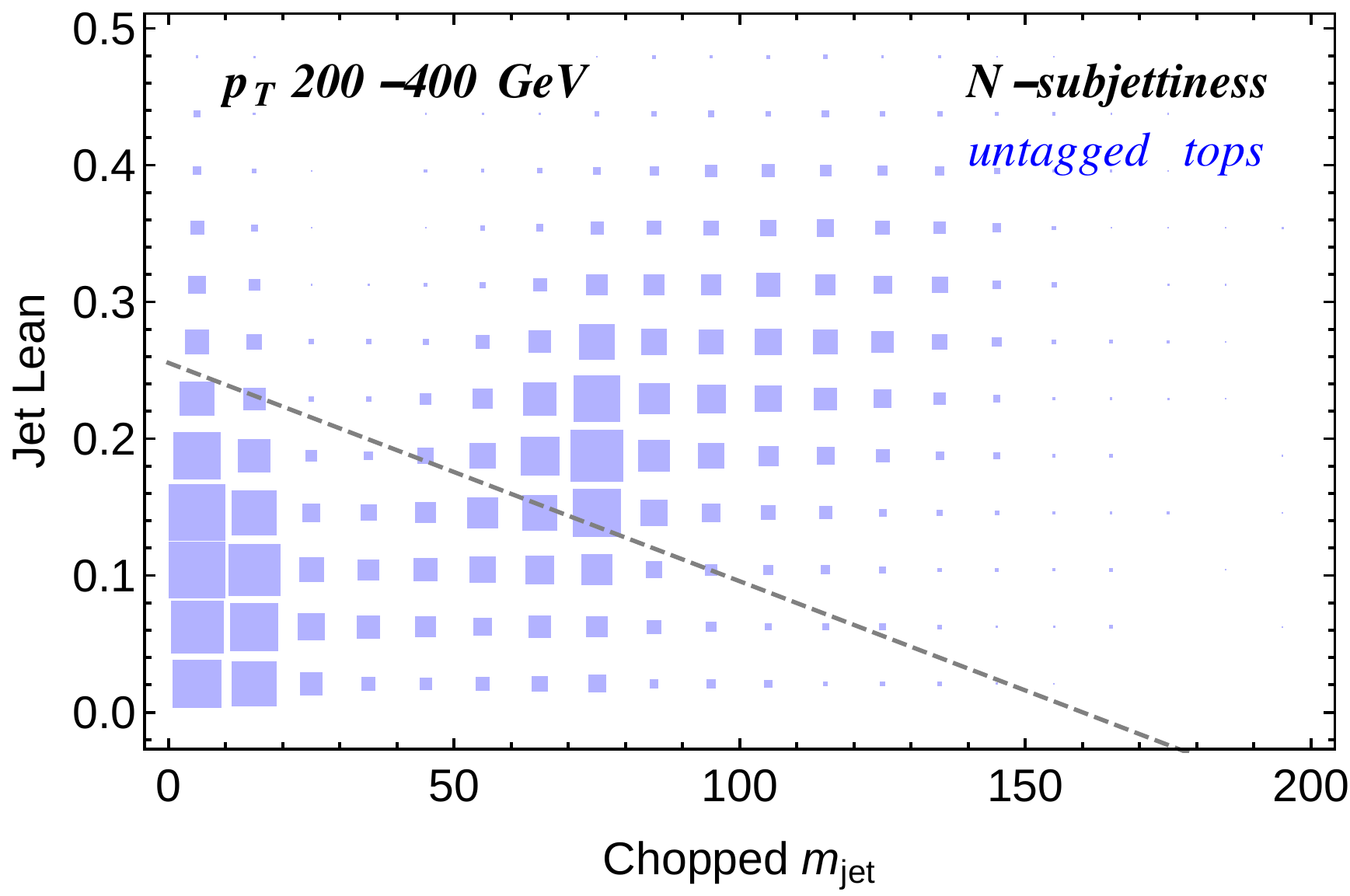} 
    \includegraphics[width=.32\textwidth]{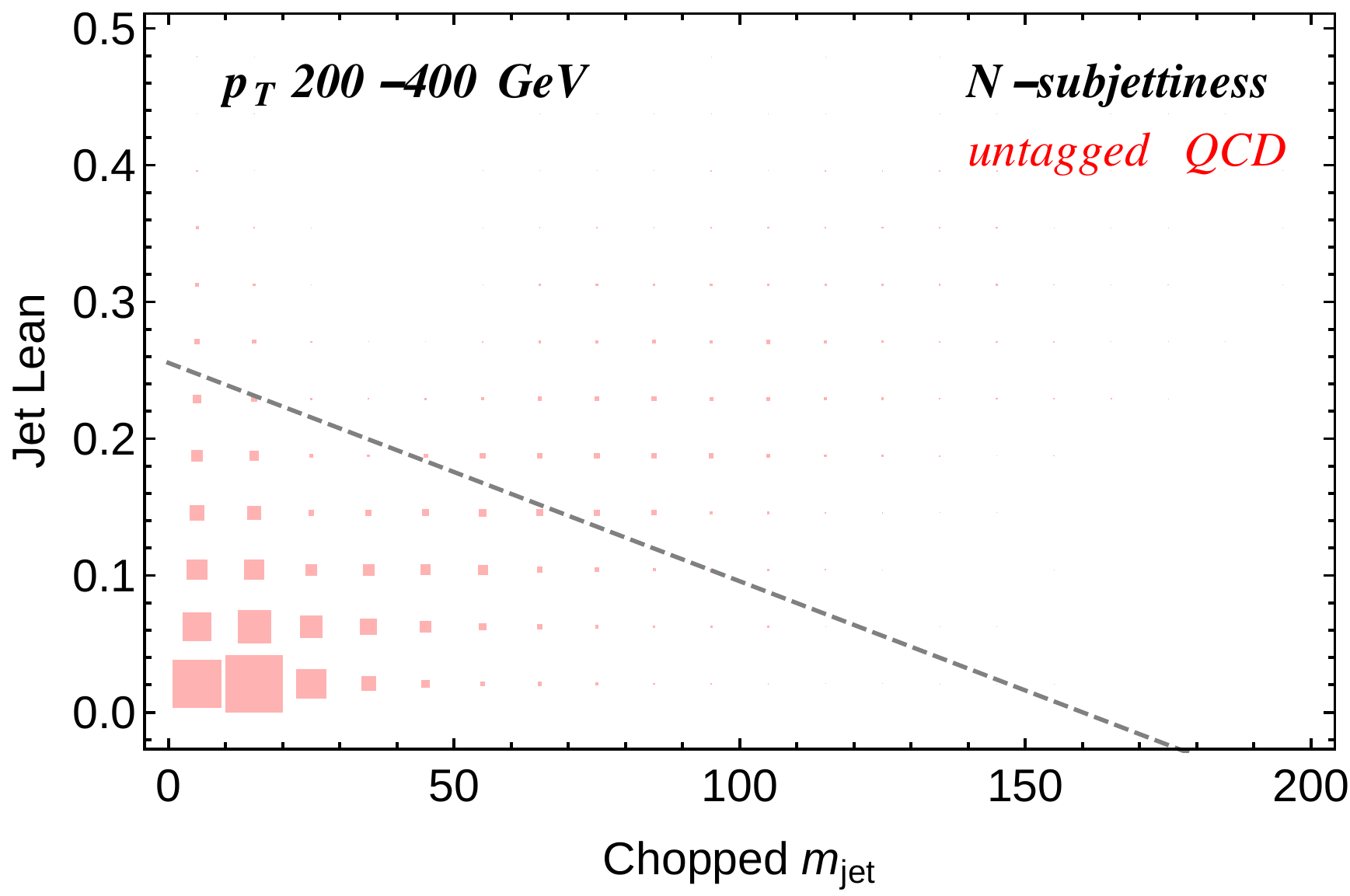} 
  \end{center} 
  \vspace{-1em}
  {\caption[1]{Two-dimensional distributions of our top tagger variables, jet lean and chopped jet mass, for top jets tagged ({\bf left}) and untagged ({\bf middle}) by $N$-subjettiness, as well as untagged QCD jets ({\bf right}). For each bin in the two-dimensional distribution, the area is proportional to the number of events in the bin.  A dashed line on each plot indicates the linear cut for the 50\% efficiency point  for the combined tagger with $N$-subjettiness.}
  \label{fig:dRtau1vsmtrim}} }
\end{figure*}

\begin{figure*}[htb!]{
  \begin{center}
    \includegraphics[width=.48\textwidth]{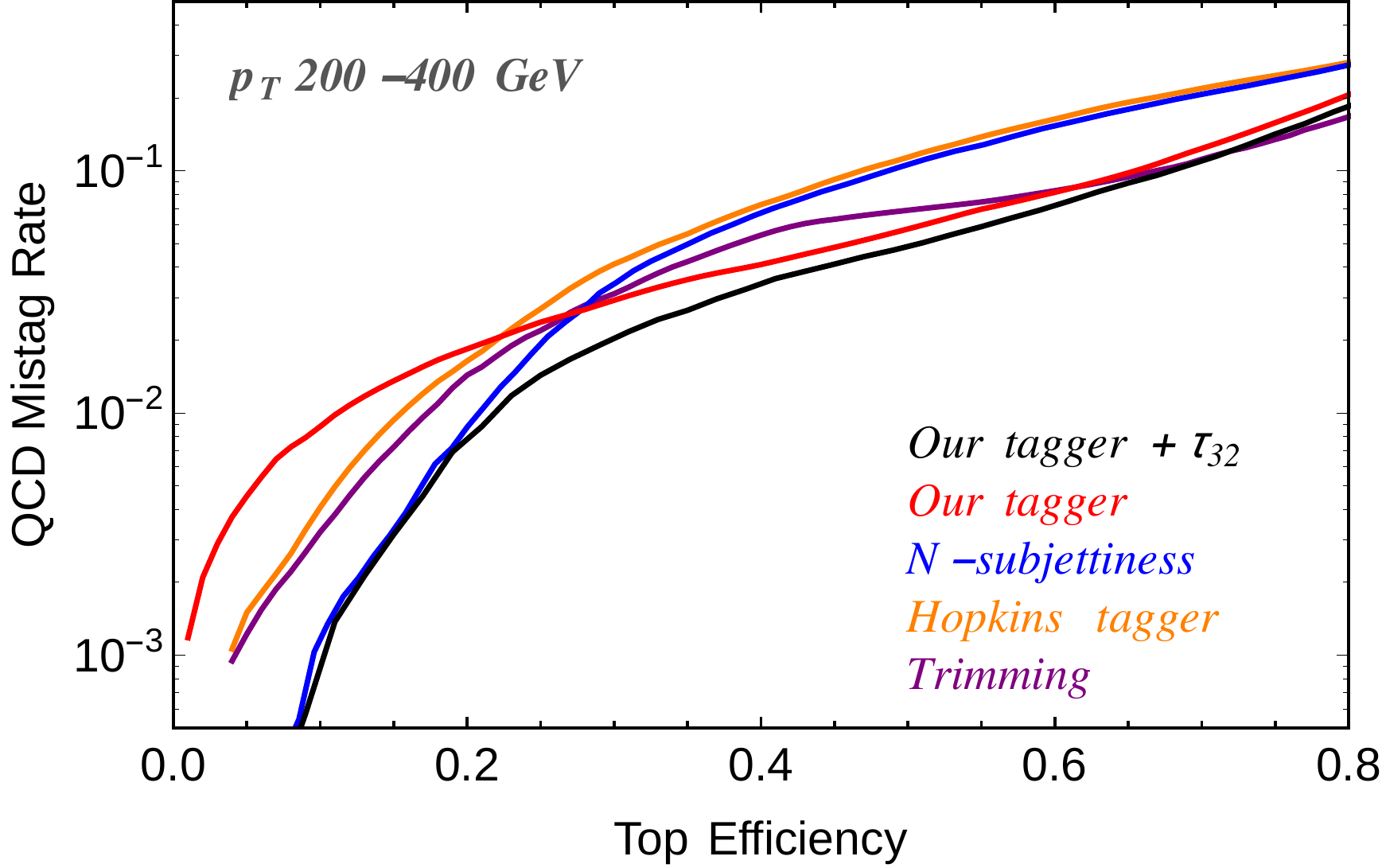}
    \includegraphics[width=.48\textwidth]{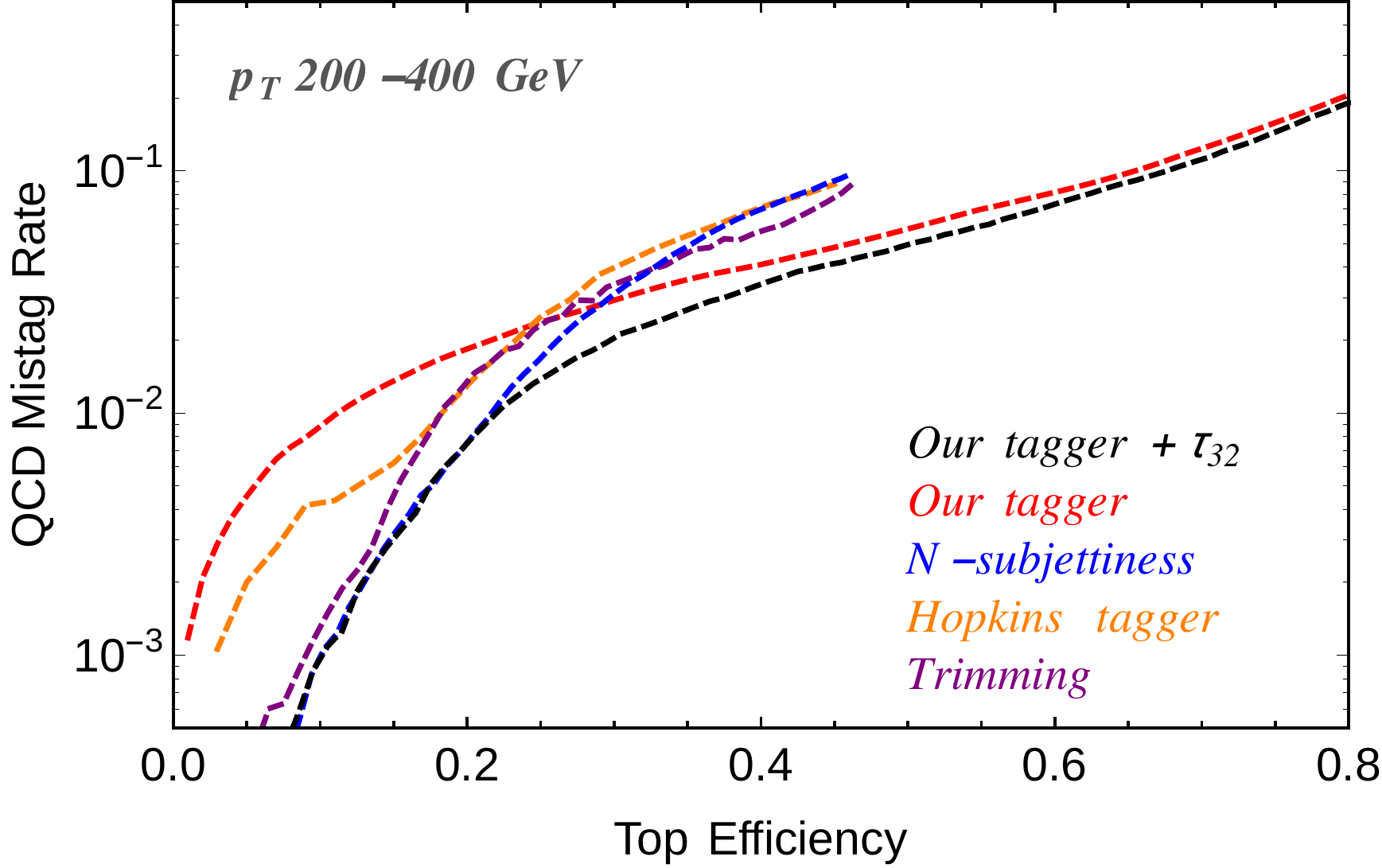}
  \end{center} 
  \vspace{-1em}
  { \caption[1]{Plots of top tagger performance in the intermediate $p_T$ range of 200-400 GeV.  The data is taken from the boosted top benchmark samples generated for the BOOST2010 workshop and the performance is shown for the trimming tagger~\cite{Krohn:2009th} (purple), Johns Hopkins tagger~\cite{Kaplan:2008ie} (orange), N-subjettiness tagger~\cite{Thaler:2011gf} (blue), the tagger studied in this paper (red) and the tagger of this paper in conjunction with the cut on $\tau_3/\tau_2$ (black). {\bf Left}: taggers with all cuts optimized for the subsample.  {\bf Right}: same as left but with an additional hard cut of $m_\text{jet} > 120\GeV$ on all jets for all taggers other than our own.}
  \label{fig:performance}} }
\end{figure*}

We wish to characterize the behavior of our top tagger in the intermediate $p_T$ range, $200\GeV < p_T < 400\GeV$, as this is where conventional top taggers start to lose efficiency. The effect of the two cuts separately is displayed in \fig{dRtau1andmtrim}, and together in \fig{dRtau1vsmtrim}, for QCD and tops that have passed and failed the conventional implementation of the $N$-subjettiness top tagger.  \fig{dRtau1vsmtrim} also displays the cut corresponding to a linear optimization at the 50\% top efficiency point. {The optimal linear Fisher discriminant~\cite{AndersonBahadur} is constructed from the two variables, while scanning over the input parameters controlling calculation of the chopped mass. As noted above, \fig{dRtau1andmtrim} shows that the distribution of both conventionally tagged and untagged tops looks markedly different from the majority of QCD events, demonstrating that the variables presented here are sensitive to discriminating structure that other taggers miss.  We use the boosted top benchmark Monte Carlo samples generated for the BOOST2010 workshop to perform our analysis.

In \fig{performance}, we show the efficiency of the top tagger described above in the intermediate $p_T$ range, along with corresponding performance from several different types of conventional taggers. In all cases, optimized cuts are found following the procedure described above, with a cut on the appropriate linear combination of output variables of the tagger used, while input parameters (where present) are scanned over a reasonable range. We carry out the above in two different cases. Initially, we do not employ the conventional hard lower cut on the total jet mass, designed to remove isolated boosted $W$-jets from the top sample. We then compare this with the imposition of an $m_J > 120\GeV$ cut on all conventional taggers. The performance in both cases is comparable with the exception of the trimming tagger at high purity, but in the second case conventional taggers cannot achieve an efficiency better than $\approx .46$ due to the hard cut. We see that unless one desires a high purity sample, our tagger on its own outperforms the conventional approaches for such a moderately boosted sample.

While {preferentially tagging poorly reconstructed tops} on its own might be of formal interest, in practical applications it would be rare to wish to tag tops only in the intermediate-boost regime. Instead, the hope would be to recover some untagged tops in addition to those already tagged by more conventional methods. Since we use the $N$-subjettiness variable as part of our method, it is natural to look at the performance of running our tagger in parallel with the more conventional cut on $\tau_3/\tau_2$, which for the 50\% efficiency point corresponds to the cut,
\begin{equation}
  \frac{\tau_3}{\tau_2} + 1.1 \left(\frac{\Delta m}{m}\right)_+
    - 1.8 \left(\frac{\Delta m}{m}\right)_- \lesssim 0.7\,,
\end{equation}
with $(\Delta m/m)_\pm$ being the fractional deviation above (below) the top mass for the given jet}.  Thus, we also show the efficiency for a double tagger for which events which pass either the conventional $\tau_3/\tau_2$ cut or our new cuts are considered tagged. It turns out that a substantial overlap in QCD mistags in both methods is present. This allows for notably better tagging performance than either approach alone, particularly in the $0.25$-$0.4$ efficiency range. In fact, the combined tagger shows at least a 50\% improvement {in QCD mistag reduction} over previously available methods in the entire $0.25$--$0.5$ efficiency range.

\section{Discussion}
\label{sec:conc}

In this note we have presented a method for differentiating (partially) boosted hadronically decaying objects without having to specify a particular topology or mass window for grouping of subjets.  Our approach allows us to extend the range of scenarios in which substructure methods can be applied, going  beyond conventional approaches that excel when all decay products are cleanly identifiable.  As an example, we considered hadronically decaying top quarks with intermediate boosts.   Here, our method alone proves more powerful 
when larger top efficiencies are required, and shows even more significant improvement when combined with a conventional top tagging method ($N$-subjettiness). Additional applications, which we leave to future work, could involve reconstruction of objects decaying both to hadrons and invisible products, for which conventional substructure approaches are entirely inapplicable.


\acknowledgments
{\bf Acknowledgments.}  We thank David Curtin, Gilly Elor, Matt Schwartz, Jesse Thaler, Jay Wacker for useful discussions. MF thanks Tel-Aviv University for its hospitality while this work was in preparation.
The work of TV is supported by the US-Israel Binational Science Foundation, the EU-FP7 Marie Curie, CIG fellowship and by the I-CORE Program of the Planning Budgeting Committee and the Israel Science Foundation (Grant No. 1937/12). The work of JRW is supported by the Office of Science, Office of High Energy Physics, of the U.S.~Department of Energy (DOE) under contract DE-AC02-05CH11231. The work of MF is supported
by the DOE under grant DE-SC003916 and the National Science Foundation (NSF) under grant No. PHY-1258729.


\end{document}